\documentclass[12pt]{article}
\usepackage{graphicx,epsfig}
\usepackage{amsfonts}
\usepackage{latexsym}
\usepackage{multirow}
\usepackage{chngpage}
\def\be{\begin{equation}}
\def\ee{\end{equation}}
\def\bea{\begin{eqnarray}}
\def\eea{\end{eqnarray}}
\newcommand{\rs}{\triangle_{\cal{R}}^{4}}
\renewcommand{\H}{{\cal H}}
\renewcommand{\k}{{\bf k}}
\newcommand{\x}{{\bf x}}
\newcommand{\f}{{\nu}}

\begin{document}

\title{{\bf Gravitational waves from an early matter era}}

\author{Hooshyar Assadullahi and David Wands\\Institute of Cosmology and Gravitation, University of Portsmouth,\\Dennis Sciama Building, Burnaby
Road, Portsmouth PO1 3FX,\\United Kingdom}
 \maketitle

\begin{abstract}
We investigate the generation of gravitational waves due to the
gravitational instability of primordial density perturbations in an
early matter-dominated era which could be detectable by experiments
such as LIGO and LISA. We use relativistic perturbation theory to
give analytic estimates of the tensor perturbations generated at
second order by linear density perturbations. We find that large
enhancement factors with respect to the naive second-order estimate
are possible due to the growth of density perturbations on
sub-Hubble scales. However very large enhancement factors coincide
with a breakdown of linear theory for density perturbations on small
scales.
To produce a primordial gravitational wave background that would be detectable with LIGO or LISA from density perturbations in the linear regime requires primordial comoving curvature perturbations on small scales of order 0.02 for Advanced LIGO or 0.005 for LISA, otherwise numerical calculations of the non-linear evolution on sub-Hubble scales are required. 
\end{abstract}

\section{Introduction}

Gravitational waves are a probe of the very early universe that go
beyond electromagnetic signals such as the cosmic microwave
background. Gravitons can propagate essentially un-scattered from
any energy scale below the Planck density. As a result they have
been considered as probes of violent events in the early universe,
such as bubble collisions \cite{Kamionkowski:1993fg}, preheating
after inflation
\cite{Khlebnikov:1997di,Easther:2006gt,Easther:2006vd,Dufaux:2007pt,Easther:2007vj,GarciaBellido:2007af,GarciaBellido:2007dg,Dufaux:2008dn}
or cosmic strings \cite{Allen:1991bk}.

Much of the work to date has been based on the local generation of
gravitational waves, e.g., due to time-varying quadrupole moment in
flat spacetime \cite{Weinberg1972}. But recent
\cite{Bruni:1996im,Mollerach:2003nq,Noh:2004bc,Nakamura:2006rk,Nakamura:2004rm,Ananda:2006af,Baumann:2007zm,Malik:2008im}
work has developed cosmological perturbation theory to deal with the
generation of gravitational waves at second order from first-order
density perturbations. Such a relativistic treatment is required for
inhomogeneities on or above the Hubble scale in an expanding
cosmology. It is possible to make a perturbative calculation of
the gravitational waves inevitably generated at second order from
the existence of linear density perturbations on all scales.

In the early radiation-dominated era, Ananda {\it et al}
\cite{Ananda:2006af} found that gravitational waves are produced
when density perturbations, which are overdamped at early times on
super-Hubble scales ($k/a\ll H$) come inside the Hubble scale ($k/a=H$)
and begin under-damped oscillations. The Hubble damping leads to the rapid
decay of the amplitude of sub-Hubble metric perturbations, and an
almost scale-invariant spectrum of gravitational waves is left on
sub-Hubble scales to propagate freely, redshifted by the
cosmological expansion.

Baumann {\em et al} \cite{Baumann:2007zm} extended this calculation
to follow the evolution of the gravitational waves generated at
second order through into the matter-dominated era, and ultimately
the present late-time acceleration using the numerical solution for
the evolution of linear density perturbations.
Their work showed a surprising behavior in the matter-dominated era,
where tensor metric perturbations grow on large-scales until
reaching a constant value once they come inside the Hubble scale, producing a
larger amplitude on scales close to the Hubble scale today than the
first-order gravitational waves generated by inflation
\cite{Baumann:2007zm}. Although intriguing, it seems very difficult
to detect such extremely long-wavelength tensor perturbations which
are only produced at late cosmic times
\cite{Mollerach:2003nq,Sarkar:2008ii}.

In this paper we will investigate the production of gravitational
waves in an {\em early} matter-dominated era, preceding the standard
radiation-dominated era, before primordial nucleosynthesis. A
matter-dominated era is expected to occur in the very early universe
after a period of inflation driven by overdamped scalar fields. At
the end of inflation the Hubble damping, $H$, decreases and the
scalar fields become massive ($m>H$). Oscillating massive scalar
fields have an effectively pressureless equation of state
\cite{Turner:1983he}. The decay of the oscillating fields leads to
the reheating of the universe and the start of the standard Hot Big
Bang model \cite{Kolb:1990vq}. If the decay is slow $\Gamma\ll H$ at
the end of inflation then we have an extended early matter-dominated
era.
%This is the ``old reheating'' picture which could be altered by
%non-perturbative decay of the inflaton fields (known as
%``preheating'' \cite{KLS,Bassett}).
An early matter-dominated era might also occur if weakly-coupled
massive scalar fields (or moduli) with non-zero vacuum expectation
values come to dominate over the inflaton decay products sometime
after inflation. In the curvaton scenario it is a weakly-coupled
massive field, rather than the inflaton, whose inhomogeneous
perturbations give rise to the primordial density perturbation
\cite{Linde:1996gt,Enqvist:2001zp,Lyth:2001nq,Moroi:2001ct}.
Gravitational waves generated at second-order have previously been studied in
the curvaton scenario \cite{Bartolo:2007vp}, in the limit where the
curvaton field does not dominate the energy density of the universe,
i.e., with no early matter era.

There are few constraints on an early matter era since the comoving
Hubble scale must be less that a few parsecs. The primordial density
perturbations on such small scales have long since been erased by
Silk damping, and a stochastic background of gravitational waves
might be one of the few ways we can probe the primordial power
spectrum on such small scales (though primordial black holes might
be another \cite{Carr:1974nx}).
Gravitational waves produced in an early matter era would have
wavelengths much smaller than the Hubble size at primordial
nucleosynthesis and thus, depending on the reheating temperature at
the end of the early matter era, could be directly directed by
gravitational wave detectors currently in operation or being
planned.

% We note that there has been considerable interest in modelling the
% production of gravitational waves from violent phase transition or
% preheating at the end of inflation
% \cite{KS,Easther,Kofman,GarciaBellido}. By contrast we are
% considering the gravitational waves produced by density
% perturbations on scales where the oscillating scalar fields are
% effectively pressureless and interact only gravitationally, until
% they decay to radiation.

This paper is organised as follows. In section~2 we summarise the
results we will need for the evolution of linear density
perturbations in a matter-dominated era. In section~3 we give the
evolution equation for second-order gravitational waves in a matter
dominated era and present the resulting power spectrum for tensor
metric perturbations at the end of the matter era. This leads to an
effective density of gravitational waves at the present day,
$\Omega_{GW,0}$, presented in section~4. In section~5 we discuss the
detectability of the gravitational wave background in possible early
matter eras, and we conclude in section~6.

\section{Density perturbations in an early matter era}

In this paper we will work in the longitudinal
\cite{Mukhanov:1990me} (or Poisson \cite{Bruni:1996im,Malik:2008im})
gauge where the perturbed metric is
\begin{equation}
 \label{metric}
  ds^{2}
   =
 a^{2}(\eta)[-(1+2\Phi)d^{2}\eta+[(1-2\Psi)\delta_{ij}+2F_{(i,j)}+h_{ij}]dx^{i}dx^{j}]
\end{equation}
where $\eta$ is conformal time, and the conformal Hubble rate is
$\H\equiv a'/a$ where primes denote a derivative with respect to
conformal time. $F_i$ describes transverse vector perturbations and
$h_{ij}$ describes tensor metric perturbations which are transverse
and trace-free.

Vector perturbations, $F_i$, must vanish at first order in the
absence of any vector part of the fluid 3-velocity, e.g., during a
period of inflation dominated by a scalar field. First-order tensor
perturbations propagate as a free field, but are rapidly redshifted
to large scales by an inflationary expansion. Gravitational waves on
smaller scales are produced from initial quantum vacuum
fluctuations, but their amplitude depends on the energy scale of
inflation. In the following we assume any first-order vector or
tensor metric perturbations are negligible and focus on the
second-order tensor perturbations that are produced from first-order
scalar perturbations which are known to exist on observable scales.

The scalar metric perturbations $\Phi$ and $\Psi$ are supported by
density and (curl-free) velocity perturbations of a fluid. In the
absence of anisotropic stress we require $\Phi=\Psi$
\cite{Mukhanov:1990me}, and the linear evolution equation for $\Phi$
is given by \cite{Malik:2008im}
\begin{equation}
\label{phinonad}
\Phi''+3{\cal{H}}\Phi'+(2{\cal{H}}'+{\cal{H}}^{2})\,\Phi=4\,\pi\,G\,a^{2}\,\delta\,P
\end{equation}

In a matter-dominated era with negligible pressure, $\delta P=0$, we
have $a\propto \eta^2$ and $\H=2/\eta$, so the linear evolution
equation for $\Phi$ is simply
\begin{equation}
 \Phi''+ \frac{6}{\eta}\Phi' = 0 \,,
\end{equation}
with the general solution $\Phi=C+D/\eta^5$, where $C$ and $D$ may
be spatially inhomogeneous, but are constant in time. Considering
only regular solutions at early times ($\eta\to0$) requires $D=0$
and we then have $\Phi$ constant in time on all scales.

We will often find it convenient to use a Fourier transform
\begin{equation}
 \label{fourier}
 \Phi(\x) = \frac{1}{(2\pi)^{\frac{3}{2}}} \int d^3\k\, \Phi_{\k}
\, e^{i\k.\x} \,,
\end{equation}
where the initial power spectrum of an isotropic distribution of
scalar metric perturbations will be a function of $k\equiv |\k|$
\begin{equation}
 \label{scalar power}
 \langle {\Phi}_{\k} {\Phi}_{\k'} \rangle
 = \frac{2\pi^{2}}{k^{3}} \delta^3(\k+\k') \, {\cal {P}}(k)
\end{equation}

Note that the linear comoving density perturbation grows in time on
all scales during a matter-dominated era and is given by
\cite{Peacock:1999ye,Malik:2008im}
\begin{equation}
\label{deltam} \frac{\delta\rho_m}{\rho} = \frac{2}{3\H^2} \nabla^2
\Phi \,.
\end{equation}
where $\nabla^{2}=\delta^{ij}\partial_{i}\partial_{j}$ is the
comoving spatial Laplacian. Because we have assumed pressure is
negligible, gravitational instability leads to a growing density
perturbation on all scales.
Eventually this will lead to a breakdown of the linear evolution on
sub-Hubble scales when the density perturbations become of order
one, corresponding to
 \begin{equation}
\label{kNL}
 k_{NL}^2(\eta) \sim {\cal P}^{-1/2}{\cal H}^2 \gg {\cal H}^2 \,.
 \end{equation}

The above is an idealisation of any realistic model. In particular
an oscillating massive scalar field with mass $m>H$ has a Compton
wavelength $\lambda\sim m^{-1}$ and hence can only be described as
pressureless matter on comoving scales $k\ll am$. Fourier modes with
$k>am$ correspond to relativistic modes with non-negligible pressure
and we expect $\Phi$ to be suppressed on these scales.
For simplicity, and to remove any dependence on the preceding
cosmological evolution, we will assume in our calculations that
there are no density perturbations on sub-Hubble scales at the
start of the early matter era
 \begin{equation}
 {\cal{P}}(k) = 0 \quad {\rm for}\ k>k_{\rm dom}
 \end{equation}
where $k_{\rm dom}=\H_{\rm dom}<am$ is the comoving Hubble scale at
the start of the matter-dominated era\footnote{In the late universe
this corresponds to the turnover in the matter power spectrum on
scales $k>k_{\rm eq}$ where $k_{\rm eq}=\H_{\rm eq}$ is the Hubble
scale at matter-radiation equality.}. Any density perturbations on
smaller scales will provide an additional, source for gravitational
waves.

Even so, density perturbations on our smallest scale will become
nonlinear, $k_{\rm dom}>k_{NL}(\eta)$, if the matter era lasts long
enough. In what follows we will present results both extrapolating
the linear results into the nonlinear regime, and also from imposing
an abrupt cut-off on the scalar power spectrum at the non-linear
scale, such that
 \begin{equation}
 {\cal{P}}(k) = 0 \quad {\rm for}\ k>k_{{\rm cut}}
 \end{equation}
where $k_{\rm cut}={\rm min}[k_{\rm dom},k_{NL}(\eta)]$, which we expect to
provide a conservative lower bound on the amplitude of
gravitational waves generated at second-order.

On very large scales the power spectrum of the primordial scalar
perturbation is commonly approximated by a power law
\begin{equation}
\label{ns} {\cal{P}}(k) = \frac{9}{25} \triangle_{\cal{R}}^{2}
 \left( \frac{k}{k_*} \right)^{n_{s}-1}
%  \quad {\rm for}\ k<k_{\rm cut}
   \,.
\end{equation}
where the numerical factor $9/25$ comes from the relation between
scalar curvature perturbation in the longitudinal and comoving
gauges on large scales in a matter dominated era
\cite{Mukhanov:1990me}.
Observations of the cosmic microwave background (CMB) radiation
\cite{Spergel:2006hy} give the amplitude,
$\triangle_{\cal{R}}^{2}\approx 2.4\times10^{-9}$, and spectral
index, $n_s\approx0.96$, for the primordial density perturbations on
very large scales $2\pi a_0/k_*\sim100$~Mpc today. We should note
however that the primordial density perturbations could be very
different on the much smaller scales relevant for the direct
detection of gravitational waves, e.g., $2\pi a_0/k\sim 1$~km.

\section{Generation of tensor perturbations}

The existence of first-order scalar perturbations inevitably leads
to second-order vector and tensor perturbations
\cite{Mollerach:2003nq,Nakamura:2006rk}.

Analogous to Eq.~(\ref{fourier}) for scalar fields, we can give the
Fourier transform of the tensor metric perturbations
(\cite{Ananda:2006af,Baumann:2007zm})
\begin{equation}
\label{tensore fourier}
 h_{ij}(x,\eta) =
 \int\frac{d^{3}\k} {(2\pi)^{\frac{3}{2}}} e^{i\k.\x}[h_{\k}(\eta)e_{ij}(\k)+\bar{h}_{\k}\bar{e}_{ij}(\k)]
 \,,
\end{equation}
where $e^{ij}(k)$ is the tensor polarization. The two polarization
tensors $e_{ij}(k)$ and $\bar{e}_{ij}(k)$ can be calculated in terms
of the orthonormal basis:
\begin{eqnarray}
\label{polarization} \nonumber
e_{ij}(k) &=&
 \frac{1}{\sqrt{2}}[e_{i}(k)e_{j}(k)-\bar{e}_{i}(k)\bar{e}_j(k)]\\
 \bar{e}_{ij}(k) &=& \frac{1}{\sqrt{2}}[e_{i}(k)\bar{e}_j(k)+\bar{e}_{i}(k)e_{j}(k)]
\end{eqnarray}
where ${\bf e}$ and $\bar{\bf e}$ are orthonormal transverse
vectors, ${\bf e}.\k=\bar{\bf e}.\k={\bf e}.\bar{\bf e}=0$ and ${\bf
e}.{\bf e}=\bar{\bf e}.\bar{\bf e}=1$.

The power spectrum of the tensor perturbations is given by
\begin{equation}
 \label{tensor power}
 \langle h_{\k}(\eta)h_{\k'}(\eta) \rangle
 = \frac12 \frac{2\pi^{2}}{k^{3}}\delta^3(\k+\k'){\cal {P}}_{h}(k,\eta)
 \,,
\end{equation}
where the factor of $1/2$ arises since, by convention, ${\cal
{P}}_{h}$ includes the contributions from both polarisations ($\bar{h}_{\k}$ as well as $h_{\k}$).

The evolution of second-order tensor mode in (\ref{metric}) is given
by the wave equation \cite{Ananda:2006af}:
\begin{equation}
\label{h''1}
 h''_{ij}+2{\cal H}  h'_{ij}+k^{2}h_{ij}=S_{ij}^{TT}
  \,,
\end{equation}
where the $S_{ij}^{TT}$ is a transverse and trace-free source term.
If we include terms up to second order in the scalar perturbations,
then $S_{ij}^{TT}$ is the transverse and tracefree part of
\cite{Ananda:2006af,Baumann:2007zm}
\begin{eqnarray}
\label{s1} \nonumber
S_{ij} &=&
 2\Phi\partial_{i}\partial_{j}\Phi - 2\Psi\partial_{i}\partial_{j}\Phi
 + 4\Psi\partial_{i}\partial_{j}\Psi + \partial_{i}\Phi\partial_{j}\Phi
 - \partial^{i}\Phi\partial_{j}\Psi - \partial^{i}\Psi\partial_{j}\Phi
 + 3\partial^{i}\Psi\partial_{j}\Psi\\
\nonumber&&
 - \frac{4}{3(1+w){\cal H}^{2}}\partial_{i}(\Psi'+{\cal
H}\Phi)\partial_{j}(\Psi'+{\cal H}\Phi)\\
&&
 -\frac{2c_{s}^{2}}{3w{\cal H }}\,[3{\cal H}({\cal
H}\Phi-\Psi')+\nabla^{2}\Psi]\,\partial_{i}\partial_{j}(\Phi-\Psi)
%\\
\end{eqnarray}
where $w=P/\rho$ is the equation of state and $c_s^2=P'/\rho'$ is
the adiabatic sound speed.

In the approximation of a pressureless matter-dominated era we have
$w=c_s^2=0$ and $\Phi=\Psi=C(\x)$ and the source term simplifies
considerably.
If we substitute Eq.~(\ref{fourier}) into Eq.~(\ref{s1}) we obtain
\begin{equation}
\label{s2} S_{ij}(x) = \frac{1}{(2\pi)^{3}} \int
 d^{3}\widetilde{\k}\,d^{3}\widetilde{\k'}\,
 [-4\widetilde{k'}_{i}\widetilde{k'}_{j}-\frac{2}{3}\widetilde{k'}_{i}\widetilde{k}_{j}]\,
 \Phi_{\widetilde{\k}}\,\Phi_{\widetilde{\k'}}\,
 e^{i(\widetilde{\k}+\widetilde{\k'}).\x}
\end{equation}

Substituting Eqs.~(\ref{s2}) and (\ref{tensore fourier}) in
Eq.~(\ref{h''1}) we find the evolution of the amplitude of each
tensor mode in Fourier space during the matter era
\begin{equation}
\label{h''fourier}
 h''_{\k} + \frac{4}{\eta} h'_{\k} + k^{2}h_{\k} = S_\k \,,
\end{equation}
where
\begin{equation}
\label{s3}
% S(k)=\frac{40}{3}(2\pi)^{-\frac{3}{2}}\,\int
% d^{3}\widetilde{k}\,e^{ij}(k)\,\widetilde{k}_{i}\,\widetilde{k}_{j}\,\Psi_{k}\,\Psi_{k-\widetilde{k}}
% \end{equation}
% The source term (\ref{s3}) may be written in the following form:
% \begin{equation}
\label{s4}
 S_\k= \frac{40}{3}{(2\pi)^{-\frac{3}{2}}}\int
d^{3}\widetilde{\k} \,e(\k,\widetilde{\k})
\,{\Phi}_{\k-\widetilde{\k}}\,{\Phi}_{\widetilde{\k}} \,,
\end{equation}
and
\begin{equation}
\label{ekk'} e(\k,\widetilde{\k})
 = e^{ij}(\k)\widetilde{k}_{i}\widetilde{k}_{j} = \widetilde{k}^{2} \sin^2\theta \,,
% [1-\mu^{2}]
\end{equation}
where $\theta$ is the angle between $\k$ and $\widetilde{\k}$,
\begin{equation}
\label{mu}
% \mu=  =
\cos\theta
 = \frac{\k.\widetilde{\k}}{k\widetilde{k}} \,.
\end{equation}

\subsection{Evolution of the tensor mode}

A striking feature of a matter dominated era is that the source term
$S_\k$ in the wave equation (\ref{h''fourier}) is constant for
linear density perturbations. This contrasts with, for example, a
radiation dominated era, where $\Phi$, and hence the source for
tensor modes, decays on sub-Hubble scales \cite{Ananda:2006af}.

The general solution of the evolution equation (\ref{h''fourier}) in the
matter era is thus
\begin{eqnarray}
\label{h1} h_\k &=&
\frac{S_\k}{k^{2}} + \sqrt{\frac{\pi}{2k^3\eta^3}}
 \left[ C_\k J_{\frac{3}{2}}(k\eta) + D_\k Y_{\frac{3}{2}}(k\eta) \right] \nonumber\\
 &=& \frac{S_\k}{k^{2}}
 + C_\k \left( \frac{\sin(k\eta)-k\eta\cos(k\eta)}{k^3\eta^{3}} \right)
 - D_\k \left( \frac{\cos(k\eta)+k\eta \sin(k\eta)}{k^3\eta^{3}} \right)
\end{eqnarray}
where $J$ and $Y$ are the Bessel functions of the first and second
kind and $C_\k$ and $D_\k$ are constants of integration.

The $J$ and $Y$ modes describe the damped, but source-free
oscillations of the gravitational field and thus have the same form
as the usual solution for first-order gravitational waves. The $Y$
mode is singular as $\eta\to0$ and rapidly decays at late times. $J$
is regular at early times and gives oscillations whose amplitude
redshifts with the expansion of the universe, $|h_\k|\propto
a^{-1}$. However the constant second-order source term, $S$,
supports a constant tensor part of the metric perturbation at late
times. This behaviour is quite different from that usually
associated with gravitational waves, reflecting the fact that this
is no longer a freely propagating wave, but rather a metric
distortion sourced by terms quadratic in first-order scalar
perturbations.

If we impose the initial condition $h=h'=0$ when $\eta=0$, then $C_\k=-3S_\k/k^2$ and the singular term is absent, $D_\k=0$, so we have the particular solution
\begin{equation}
\label{h2}
 h_\k =
 \frac{S_\k}{k^{2}} \left[ 1+3 \left( \frac{k\eta\cos(k\eta)-\sin(k\eta)}{k^{3}\eta^{3}} \right) \right]
% =\frac{2}{5}\frac{S}{\H^2} f(k\eta)
 \,.
\end{equation}
At early times, or equivalently in the large scale, super-Hubble limit $(k\ll{\cal H)}$, we find a growing tensor perturbation
\begin{equation}
\label{h3}
% f(k\eta)=1 \Leftrightarrow
 h_\k = \frac{S_\k}{10} \eta^2
% = \frac{2}{5}\frac{S}{{\cal H} ^{2}}
 \,.
\end{equation}
We see that the large-scale tensor mode grows at the same rate as
the comoving density perturbation (\ref{deltam}) in the matter era
%
%From the Poisson equation we have :
%\begin{equation}
%\frac{\nabla^{2}}{a^{2}}\Phi=4\pi G \delta
%\rho_{(c)}
%\end{equation}
%Where $\delta\rho_{c}$ denotes the density perturbation in the
%comoving gauge. Thus we have:
\begin{equation}
\label{hdeltarho}
 h \propto \frac{\delta\rho_m}{\rho} \propto \frac{1}{{\cal H}^{2}}
 \,.
\end{equation}
The tensor amplitude grows until the mode enters the Hubble scale and
at late times, or on sub-Hubble scales $(k\gg{\cal H})$, it becomes constant
\begin{equation}
\label{h3 small scale}
% f(k\eta)=\frac{5}{2}\frac{{\cal H}^{2}}{k^{2}}\Leftrightarrow
h_\k = \frac{S_\k}{k^{2}}
\end{equation}

Substituting Eq.~(\ref{s4}) for $S_\k$ into Eq.~(\ref{h2}) we will
write the solution for the tensor perturbation as
\begin{equation}
\label{h4}
 h_\k
% =
% \frac{16\times{(2\pi)^{-\frac{3}{2}}}}{3{\cal H}^{2}}
%  f(k\eta)\times\int^{k_{dom}}_{0}d^{3}\widetilde{k} e(k,\widetilde{k})
% \hat{\phi}_{k-\widetilde{k}}\hat{\phi}_{\widetilde{k}}
 =
 \frac{40 g(k\eta)}{3{(2\pi)^{3/2}}}
 k^{-2} \int^{k_{dom}}_{0}d^{3}\widetilde{\k} \, e(\k,\widetilde{\k})
 \, {\Phi}_{\k-\widetilde{\k}}{\Phi}_{\widetilde{\k}}
 \,,
\end{equation}
where the growth function for the tensor modes is given by
\begin{equation}
 \label{defg}
 g(k\eta) = 1+3 \left( \frac{k\eta\cos(k\eta)-\sin(k\eta)}{k^{3}\eta^{3}} \right) \,,
\end{equation}
which approaches unity at late times on sub-Hubble scales.

% [PREVIOUSLY USED $f(k\eta)=(5\H^2/2k^2)g(k\eta)=(10/k^2\eta^2)g(k\eta)$ WHICH WAS UNITY AT EARLY TIMES/LARGE SCALES AND APPROACHED $10/k^2\eta^2$ AT LATE TIMES/SMALL SCALES.]

\subsection{The power spectrum of the gravitational waves}

{}From (\ref{h4}) we can immediately write down the two-point
function for the tensor modes
\begin{equation}
\label{hh'1} \langle h_{\k}(\eta)h_{\k'}(\eta)\rangle = \left(
\frac{40g(k\eta)}{3{(2\pi)^{3/2}}} \right)^{2}
 k^{-4} \int^{\k_{dom}}_{0} \, d^{3}\widetilde{\k}\, e(\k,\widetilde{\k})
\,d^{3}\widetilde{\k'}\, e(\k',\widetilde{\k'}) \,
\langle\Phi_{\k-\widetilde{\k}}\,\Phi_{\widetilde{\k}}\,\Phi_{\k'-\widetilde{\k'}}\,\Phi_{\widetilde{\k'}}\rangle
\end{equation}
where on the right-hand-side for a Gaussian distribution of scalar
perturbations we have (for non-zero $k$ and $k'$)
\begin{eqnarray}
\label{phi4}
\langle\Phi_{\k-\widetilde{\k}}\,\Phi_{\widetilde{\k}}\,\Phi_{\k'-\widetilde{\k}'}\,\Phi_{\widetilde{\k}'}\rangle
=
% first term automatically zero for non-zero k
% \langle\Phi_{\k-\widetilde{\k}}\,\Phi_{\widetilde{\k}}\rangle\langle \Phi_{\k'-\widetilde{\k'}}\,\Phi_{\widetilde{\k'}}\rangle+&&
% \nonumber\\
\langle\Phi_{\k-\widetilde{\k}}\,\Phi_{\k'-\widetilde{\k}'}\rangle\langle\Phi_{\widetilde{\k}}\,\Phi_{\widetilde{\k}'}\rangle+
\langle\Phi_{\k-\widetilde{\k}}\,\Phi_{\widetilde{\k}'}\rangle\langle\Phi_{\widetilde{\k}}\,\Phi_{\k'-\widetilde{\k}'}\rangle
\end{eqnarray}
\\
% The first term on the right hand side of the equation (\ref{phi4})
% is zero for $k\neq0$.
% The two remaining terms are given by
% Eq.~(\ref{scalar power}) as
% \begin{equation}
% \label{phi22}
% \langle\Phi_{\k-\widetilde{\k}}\,\Phi_{\k'-\widetilde{\k'}}\rangle\langle\Phi_{\widetilde{\k}}\,\Phi_{\widetilde{\k'}}\rangle=\frac{2\pi^{2}}{(k-\widetilde{k})^{3}}{\cal{P}}(\k-\widetilde{\k})\,\delta^{3}{\bf
% {(\k+\k'-\widetilde{\k}}-\widetilde{\k'})}
% \frac{2\pi^{2}}{\widetilde{k}^{3}}\,{\cal{P}}(\widetilde{\k})\,\delta^{3}{\bf(\widetilde{\k}+\widetilde{\k'})}
% \end{equation}
% and
% \begin{equation}
% \label{phi22'}
% \langle\Phi_{\k-\widetilde{\k}}\,\Phi_{\widetilde{\k'}}\rangle\langle\Phi_{\widetilde{\k}}\,\Phi_{\k'-\widetilde{\k'}}\rangle=\frac{2\pi^{2}}{(\widetilde{k'})^{3}}\,{\cal{P}}(\widetilde{\k'})\,\delta^{3}{\bf{(\k+\widetilde{\k'}-\widetilde{\k})}}
% \frac{2\pi^{2}}{\widetilde{k}^{3}}\,{\cal{P}}(\widetilde{\k})\,\delta^{3}\bf{(\widetilde{\k}+\k'-\widetilde{\k'})}
% \end{equation}
Substituting Eqs.~(\ref{phi4}) and (\ref{scalar power}) into
Eq.~(\ref{hh'1}) we obtain
\begin{eqnarray}
\label{hh'2} \nonumber\langle h_{\k}(\eta)h_{\k'}(\eta)\rangle &=&
 \left( \frac{40 g(k\eta)}{3} \right)^{2}\,
 \frac{\pi\delta^{3}(\k+\k') }{2k^{4}} \, \\
&& \quad \times
 \int d^{3}\widetilde{\k} \,\,
 e(\k,\widetilde{\k}) \left[ e(\k',\widetilde{\k}) +
 e(\k',\k-\widetilde{\k}) \right]
 \,{\cal{P}}(\k-\widetilde{\k})\,{\cal{P}}\,(\widetilde{\k})
 \,,
% \\
%
% &=&
%  \left( \frac{40 g(k\eta)}{3(2\pi)^{3/2}} \right)^{2}\, k^{-4}
%  \int^{k_{dom}}_{0}d^{3}\widetilde{\k} \,\, \widetilde{k}^{2}\,
%  \,\frac{4\pi^{4}(1-\mu^{2}}{\widetilde{k}^3|\k-\widetilde{\k}|^{3}}\,{\cal{P}}(\k-\widetilde{\k})\,{\cal{P}}\,(\widetilde{\k})
%  \\
% &&
% \times[\widetilde{k}^{2}(1-(\frac{\k.\widetilde{\k}}{k\widetilde{k}})^{2}+\widetilde{\widetilde{k}}^{2}(1-(\frac{\k.\widetilde{\widetilde{\k}}}{k\widetilde{\widetilde{k}}})^{2})]\delta^{3}{\bf(\k+\k')}
%  \,,
\end{eqnarray}
where
% \begin{equation}
% \label{ktiltil}\bf{ \widetilde{\widetilde{\k}}=\k-\widetilde{\k}}
%  \,.
% \end{equation}
$e({\bf p},{\bf q})=q^2\sin^2\theta$ is defined in Eq.~(\ref{ekk'}).

% If we substitute (\ref{ns}) in ({\ref{hh'2}) we get:
% \begin{eqnarray}
% \label{hh'2'} \nonumber\langle
% h_{k}(\eta)h_{k'}(\eta)\rangle&=&10\times
% f^{2}(k\eta)\times\frac{\rs}{{\cal{H}}^{4}}\int^{k_{dom}}_{0}d^{3}\widetilde{k}\widetilde{k}^{2}(1-\mu^{2})
% \frac{1}{\widetilde{k}^{3}(k-\widetilde{k})^{3}}(\frac{k-\widetilde{k}}{k_{cmb}})^{n_{s}-1}(\frac{\widetilde{k}}{k_{cmb}})^{n_{s}-1}\\
%&\times&[\widetilde{k}^{2}(1-(\frac{k.\widetilde{k}}{k\widetilde{k}})^{2}+\widetilde{\widetilde{k}}^{2}(1-(\frac{k.\widetilde{\widetilde{k}}}{k\widetilde{\widetilde{k}}})^{2})]\,\delta^{3}{\bf(k+k')}
%\end{eqnarray}

For simplicity we will assume that the power spectrum of the
primordial scalar perturbations given in Eq.~(\ref{ns}) is
effectively scale invariant $n_s=1$ on scales $k<k_{\rm dom}$.
%\begin{equation}
%\label{ns=1} n_{s}\approx 1
%\end{equation}
%If we substitute (\ref{hh'2'}) in (\ref{ns=1}) we can generate the following:
%
The power spectrum (\ref{tensor power}) of gravitational waves
generated is then
% \begin{eqnarray}
% \label{hh'3}\nonumber\langle h_{\k}(\eta)h_{\k'}(\eta)\rangle
%  &=&
% % 10\times f^{2}(k\eta) \times \frac{\rs}{{\cal{H}}^{4}}
%  32\pi g^2(k\eta) \rs k^{-4} \\
% && \times
% \int^{k_{dom}}_{0}d^{3}\widetilde{\k}\widetilde{k}^{2}(1-\mu^{2})
%  \frac{1}{\widetilde{k}^{3}(k-\widetilde{k})^{3}}\\
% && \times
% [\widetilde{k}^{2}(1-(\frac{\k.\widetilde{\k}}{k\widetilde{k}})^{2}+\widetilde{\widetilde{k}}^{2}(1-(\frac{\k.\widetilde{\widetilde{\k}}}{k\widetilde{\widetilde{k}}})^{2})]\,\delta^{3}{\bf(k+k')}
% \end{eqnarray}
% With considering the definition of the tensor power spectrum
% (\ref{tensor power}) we have
% \begin{eqnarray}
% \label{p1} \nonumber{\cal{P}}_{h}(k,\eta)&=&
% % \frac{1}{2}\times\frac{\rs}{{\cal H}^{4}} \times f^{2}(k\eta)\times k^{3}\\
%   \frac{144}{25\pi} g^2(k\eta) \rs k^{-1} \\
% && \times \int^{k_{dom}}_{0}
% \frac{d^{3}\widetilde{\k}(1-\mu^{2})}{\widetilde{k}(k-\widetilde{k})^{3}}\,[\widetilde{k}^{2}(1-(\frac{\k.\widetilde{\k}}{k\widetilde{k}})^{2}+\widetilde{\widetilde{k}}^{2}(1-(\frac{\k.\widetilde{\widetilde{\k}}}{k\widetilde{\widetilde{k}}})^{2})]
% \end{eqnarray}
%If we change the integral variable we get:
\begin{equation}
\label{p2} {\cal{P}}_{h}(k,\eta)
 =
% \frac{1}{2}\times\frac{\rs}{{\cal H}^{4}}\times f^{2}(k\eta)\times k^{3} \times(2\pi\times k_{dom})
 2 \left( \frac{24 g(k\eta)}{5} \right)^2 \rs \left( \frac{k_{\rm dom}}{k} \right) I_{1}(k/k_{dom})
 \,,
\end{equation}
where the integral
\begin{equation}
I_{1}(k/k_{\rm dom})
 = \frac{1}{2\pi k_{\rm dom}} \int d^3\widetilde{\k}\,
 \frac{[e(\k,\widetilde{\k})]^2}{\widetilde{k}^3|\k-\widetilde{\k}|^3}
  \, \Theta(k_{\rm dom}-\widetilde{k}) \, \Theta(k_{\rm dom}-|\k-\widetilde{\k}|) \,,
\end{equation}
can be written as
\begin{equation}
\label{i1}  I_{1}(x)= \int_{-1}^{1}d\mu
 \int^{1}_{0} dy \frac{(1-\mu^{2})^{2} y^{3}}{(x^{2}+y^{2}-2xy\mu)^{\frac{3}{2}}}
  \, \Theta(1-x^2-y^2+2xy\mu) \,,
\end{equation}
and we introduce the Heaviside step function, $\Theta(k_{\rm
dom}-q)$, to cut-off the scalar spectrum, ${\cal P}(q)$, on small
scales $q>k_{\rm dom}$.

% and
% \begin{equation}
% \label{i2} I_{2}=\int^{-1}_{1}\,d\mu\,\int^{1}_{0}
% (\frac{y(1-\mu^{2})}{(x^{2}+y^{2}-2xy\mu)^{\frac{1}{2}}})(1-\frac{(x-y\mu)^{2}}{(x^{2}+y^{2}+2xy\mu)})
% \,dy
% \end{equation}
% and we introduce
% \begin{eqnarray}
% \label{x} x=\frac{k}{k_{dom}}\\
% \label{y} y=\frac{\widetilde{k}}{k_{dom}}
% \end{eqnarray}

In the limit $x\ll 1$ the step function is equal to one throughout
the integral in Eq.~(\ref{i1}) and we obtain the analytic expression
\begin{equation}
\label{x} I_{1}(x) \approx \frac{16}{15} - \frac{4}{3}x +
\frac{16}{35}x^{2} \,.
\end{equation}
For any $x<1$ we can take $I_1$ to be a numerical factor
$\leq16/15$.

% [PRESENT GRAPH OF THE INTEGRAL, $I_1(x)$, EVALUATED NUMERICALLY
% INCLUDING THE STEP FUNCTION.]
%\begin{center}
 % {\includegraphics[width=8 cm,  angle=270]{plot}}
 %\end{center}
 %\caption{Evolution of $I_{1}$ or $I_{2}$ when considering different $X=k/k_{dom}$}
%\end{figure}

\begin{figure}
\scalebox{.75} { \includegraphics*{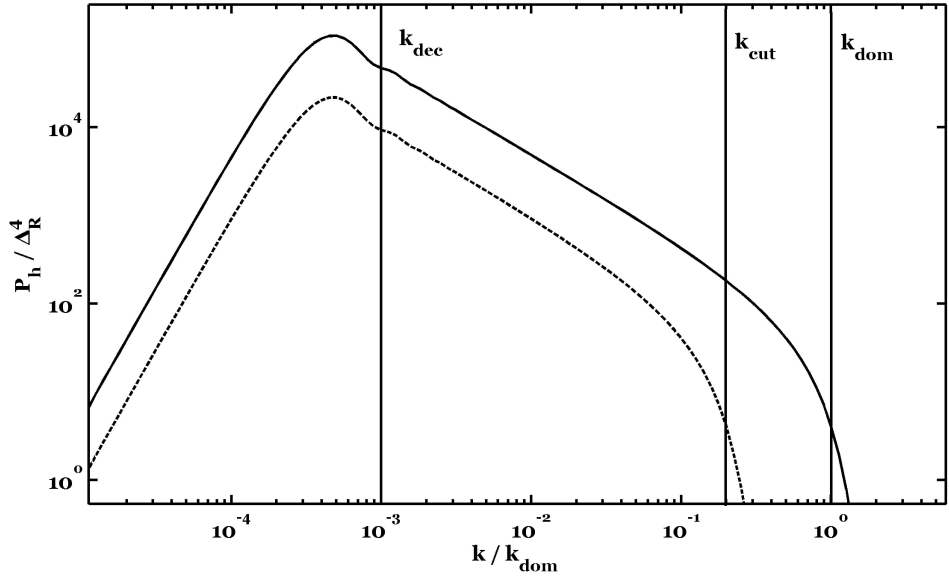} } \caption{The
power spectrum of gravitational waves, shown as a function of
wavenumber $k$, generated from scalar perturbations during a matter
dominated era, ${\cal P}_h(k,\eta_{\rm dec})$. The solid line shows
the prediction using the linear matter power spectrum down to
$k_{\rm dom}=10^3k_{\rm dec}$, the comoving Hubble scale at the
start of matter domination. The dotted line shows the prediction
when the matter power spectrum is truncated at $k_{\rm
cut}=200k_{\rm dec}$. $k_{\rm dec}$ denotes the Hubble scale at the
end of the matter era.}\label{fig:power}
\end{figure}

Equation (\ref{p2}) thus gives a simple numerical estimate of the
tensor perturbation generated at second order during a matter
dominated era. In particular at the end of an early matter-dominated
era we have
\begin{equation}
\label{p4} {\cal{P}}_{h}(k,\eta_{dec}) \simeq 2
\left(\frac{24}{5}\right)^2 \rs \times \left( \frac{k_{dom}}{k}
\right) \times g^{2}(k/k_{\rm dec}) \times I_1(k/k_{\rm dom})\,,
\end{equation}
where $k_{\rm dec}=\eta_{\rm dec}^{-1}$ is the Hubble scale at the
start of the radiation-dominated era, when the matter decays into
radiation.
This power spectrum for tensor perturbations is shown by the solid
line in Figure~\ref{fig:power} for an example where $k_{\rm
dec}=10^3 k_{\rm dom}$.

% \subsubsection{Large scales ($k\ll k_{\rm dec}$)}

In the super-Hubble limit the tensor amplitude grows and we have
$g(k\eta)\simeq k^2\eta^2/10$ from Eq.~(\ref{defg}). Thus on scales
larger than the Hubble size at the end of the matter era ($k<k_{\rm
dec}$) we have
\begin{equation}
 \label{Phdecsuper}
{\cal P}_h(k,\eta_{\rm dec}) \simeq 0.5 \rs \left( \frac{k_{\rm
dom}k^3}{k_{\rm dec}^4} \right) \,,
\end{equation}
where we have taken $I_1(k/k_{\rm dom})\simeq 16/15$ using
Eq.~(\ref{x}).
The tensor perturbations thus have a steep blue spectrum on large
scales, and are strongly suppressed on super-Hubble scales at the
end of the matter era.

% Also if we take the large scale limit, the maximum amount of
% gravitational waves comes from the mode which goes inside the
% Hubble, at the time that moduli fields decay ($k=k_{dec}$) (it is
% the biggest k that we can consider in the large scale limit).
% \\
% \\
% Hence from (\ref{p4}) we can generate the maximum amount of
% gravitational waves in large scale according to the following:
% \begin{equation}
% \label{p5}
% {\cal{P}}_{h}(k_{dom},\eta_{dec})=1.2\times(\frac{k_{dom}}{k_{dec}})\times\rs
% \end{equation}

% \subsubsection{Smaller scales ($k_{\rm dec}<k< k_{\rm dom}$)}

On scales which enter the Hubble scale during the early matter dominated
era $k_{\rm dec}<k< k_{\rm dom}$ then we find that the tensor
amplitude has settled down to a constant value by the end of the
matter era, $g(k\eta_{\rm dec})\simeq 1$ and we have
\begin{equation}
 \label{Phdecsub}
 {\cal P}_h(k,\eta_{\rm dec}) \simeq 46 \rs \left( \frac{k_{\rm
dom}}{k} \right) I_1(k/k_{\rm dom}) \,,
\end{equation}
Thus we find a decreasing power spectrum for the tensor modes on
small scales with ${\cal P}_h\propto k^{-1}$ if we take
$I_1(k/k_{\rm dom})$ to be constant. In practice $I_1$ becomes small
for $k\sim k_{\rm dom}$ leading to an additional suppression on the
smallest scales.

% If we take the small scale limit $f(k\eta)=\frac{5}{2}\frac{{\cal
% H}^{2}}{k^{2}}$ (\ref{h3 small scale}). If we Substitute it in
% (\ref{p4}) after simplification we have:
% \begin{equation}
% \label{pss} {\cal{P}}_{h}=7.5\times\rs\times(\frac{k_{dom}}{k})
% \end{equation}

Thus we find the maximum amplitude of tensor perturbations is
generated at the scale just entering the Hubble scale at the end of
the matter era, $k_{\rm dec}$, for which we have
\begin{equation}
 \label{Phmax}
 {\cal P}_h^{\rm max} = {\cal P}_h(k_{\rm dec},\eta_{\rm dec})
 \simeq 50 \rs \left( \frac{k_{\rm dom}}{k_{\rm dec}} \right) \,,
\end{equation}
where we have taken $I_1(k_{\rm dec}/k_{\rm dom})\simeq 16/15$ for
$k_{\rm dec}\ll k_{\rm dom}$ using Eq.~(\ref{x}).

Although the tensor power spectrum generated at second order is
necessarily proportional to the square of the first-order scalar
power spectrum, we find that the constant source term from scalar
perturbations, $S_{ij}$ in Eq.~(\ref{s2}), extending to sub-Hubble
scales, leads to a additional factor $k_{\rm dom}/k_{\rm dec}$ which
may be large depending upon the duration of the early matter
dominated era. Thus we find that the tensor power spectrum may be
significantly enhanced with respect to the naive estimate ${\cal
P}_h \sim \rs$.

% In above formula we consider the Hubble parameter at the time that
% moduli field decays, also this equation demonstrates that the power
% spectrum of the gravitational waves from moduli field in small
% scale, has it's largest value, for the mode that goes inside the
% Hubble at the time of field's decay, because we want to see the
% maximum amount of the gravitational waves we have;
% \begin{equation}
% \label{pssmax}
% {\cal{P}}_{h}=7.5\times\rs\times(\frac{k_{dom}}{k_{dec}})
% \end{equation}
%
% We can see that for the modes which are so small ($k>k_{dom}$) $\phi
% $ is not constant and it becomes so small:
% \begin{equation}
% \label{phismall}
% \Phi(\widetilde{k}\eta)=\frac{1}{1+(\frac{k}{k_{dom}})^{2}}\,\,\,,k>>k_{dom}\Longrightarrow\Phi(\widetilde{k}\eta)<<
% 1
% \end{equation}
% So this effect makes the gravitational waves power spectrum so
% small, which can not be detectable. For this reason we have not
% considered the very small scales.

\subsection{Nonlinear cut-off}

The gravitational wave power spectrum (\ref{Phmax}) becomes largest
when the comoving Hubble scale at the end of the matter era, $k_{\rm
dec}$, becomes much larger than the smallest scale $k_{\rm dom}$.
But as the Hubble scale grows, the comoving density contrast
(\ref{deltam}) becomes large on scales far inside the Hubble scale,
signalling a breakdown of the linear results used thus far to
estimate the source term in Eq.~(\ref{h''fourier}).

Below the non-linear scale, $k_{NL}$ defined in Eq.~(\ref{kNL}), a
perturbative analysis suggests that power will be rapidly transfered
to smaller scales \cite{KKprivate} leading to a suppression of
$\Phi_{\bf k}$ and thus the source, $S_\k$, on scales $k>k_{NL}$. A
full analysis of the nonlinear regime requires a full numerical
treatment, such as a lattice field theory calculation, as has been
performed in preheating models at the end of inflation, or an N-body
simulation in the (small-scale) Newtonian regime.

In practice we will obtain a conservative lower limit on the
amplitude of primordial gravitational waves by assuming the scalar
power spectrum is vanishing on all scales $k>k_{NL}(\eta)$ instead
of the fixed comoving cut-off, $k_{\rm dom}$.
This implies a time-dependent source term (\ref{s4}) since the upper
limit of the integral in k-space becomes time-dependent.

For $k<k_{NL}$ we find $S_\k k_{\rm NL}^2{\cal P}$ and thus
\begin{equation}
{{S}_{\bf k}}' = {{k}_{NL}}' \frac{\partial}{\partial k_{NL}} S_{\bf
k} \sim {\cal H} S_{\bf k} \,.
\end{equation}
The rate of change of the source term is thus slow compared with the
decay time for the transient part of the solution in Eq.~(\ref{h1})
for $k>\H$ and thus we take the quasi-static generalisation of
Eq.~(\ref{h3 small scale}) for sub-Hubble scales
\begin{equation}
h_\k(\eta) \simeq \frac{S_\k[k_{NL}(\eta)]}{k^2} \,.
\end{equation}

Thus a very conservative lower bound on the power spectrum of tensor
perturbations generated on sub-Hubble scales at the end of an early
matter-dominated era ($k_{\rm dec}<k<k_{\rm cut}$) is given by the
generalisation of (\ref{Phdecsub})
\begin{equation}
 \label{Phdecsub3}
 {\cal P}_h(k,\eta_{\rm dec}) \simeq 46 \rs \left( \frac{k_{\rm
cut}}{k} \right) I_1(k/k_{\rm cut}) \,,
\end{equation}
where we have $k_{\rm cut}={\rm min}\{k_{\rm dom},k_{NL}(\eta_{\rm
dec})\}$ and
\begin{equation}
 k_{NL}(\eta_{\rm dec}) \simeq {\cal P}^{-1/4} k_{\rm dec} \sim 200\, k_{\rm dec} \,.
\end{equation}

For $k\gg k_{NL}$ we find that either $\Phi_{\tilde{\k}}$ or
$\Phi_{\k-\tilde{\k}}$ in the integrand in Eq.~(\ref{s4}) vanishes
for any $\tilde{\k}$ and the source term, $S_\k$, goes to zero.
Equation~(\ref{h1}) for $S_\k=0$ reduces to the standard solution
for a free gravitational wave in a matter-dominated era, whose
amplitude is redshifted, $|h_\k|\propto a^{-1}$, on small scales,
$k>k_{NL}\gg\H$. Thus we obtain
\begin{equation}
\label{Pkcut}
 {\cal P}_h(k,\eta) \simeq 46 \rs \left( \frac{k_{NL}(\eta)}{k} \right)^{4}
\end{equation}
Hence ${\cal P}_h(k,\eta_{\rm dec})\propto k^{-4}$, and the power
spectrum of the gravitational waves is strongly suppressed on small
scales ($k\gg k_{NL}(\eta)$) if non-linear evolution suppresses the
scalar perturbation on these scales.

Assuming the scalar perturbations rapidly decay to effectively zero
on non-linear scales may be unduly pessimistic as in fact
gravitational instability continues on small scales and
non-linearity transfers power to smaller scales. However eventually
the scalar metric perturbation must decay on the smallest scales,
where the velocity of the matter becomes non-negligible. We leave a
detailed numerical calculation for future work and henceforth
present the predicted gravitational wave background using both the
linear result on all scales and the linear result cut-off at the
non-linear scale.

\section{Present density of gravitational waves}

The effective energy density of a stochastic background of
gravitational waves, on scales much smaller than the Hubble scale,
is given by \cite{Maggiore:1999vm}
\begin{equation}
\label{maggiore}
\rho_{GW} = \frac{1}{32\pi G} \langle \dot{h}_{ij} \dot{h}^{ij}
\rangle = \frac{k^2}{32\pi G a^2} \int d(\ln k)\ {\cal P}_h(k,\eta)
\,.
\end{equation}
Note that our second-order tensor modes (\ref{h2}) produced by
linear scalar perturbations become constant on sub-Hubble scales in
the early matter dominated era and therefore have negligible energy
density at that time. They behaves quite differently from the
conventional view of gravitational waves. But as the Newtonian
potential $\Phi$ decays on sub-Hubble scales either due to
non-linear evolution on small scales or in the radiation era on
sub-Hubble scales, the source term $S_{\k}$ for tensor modes also
decays on sub-Hubble scales and we are left with freely oscillating
gravitational waves.
%This is demonstrated in a simple numerical example in an appendix.

The fraction of the critical energy density in gravitational waves
per logarithmic range of wavenumber $k$ in the radiation era is
\begin{equation}
\label{omegap}
 \Omega_{GW}(k,\eta) = \frac{1}{12} \left( \frac{k}{\H} \right)^2 {\cal P}_h(k,\eta)
  \,.
\end{equation}
During and after the radiation-dominated era, the density of
gravitational waves on sub-Hubble scales then redshifts exactly as
any non-interacting relativistic particles and in the present day we
have
\begin{equation}
 \label{OmegaGW0}
 \Omega_{GW,0}(k) \simeq
  \frac{\Omega_{\gamma,0}}{12} \left( \frac{k}{k_{\rm dec}} \right)^2 {\cal P}_h(k,\eta_{\rm dec})
  \,,
\end{equation}
where the present density of photons is $\Omega_{\gamma,0}\simeq
%2.5\times10^{-5}h^{-2}
1.2\times10^{-5}$, and we neglect additional numerical factors due
to the detailed thermal history, such as the heating of photons by
the annihilation of other relativistic particle species.

\begin{figure}
\scalebox{.75} { \includegraphics*{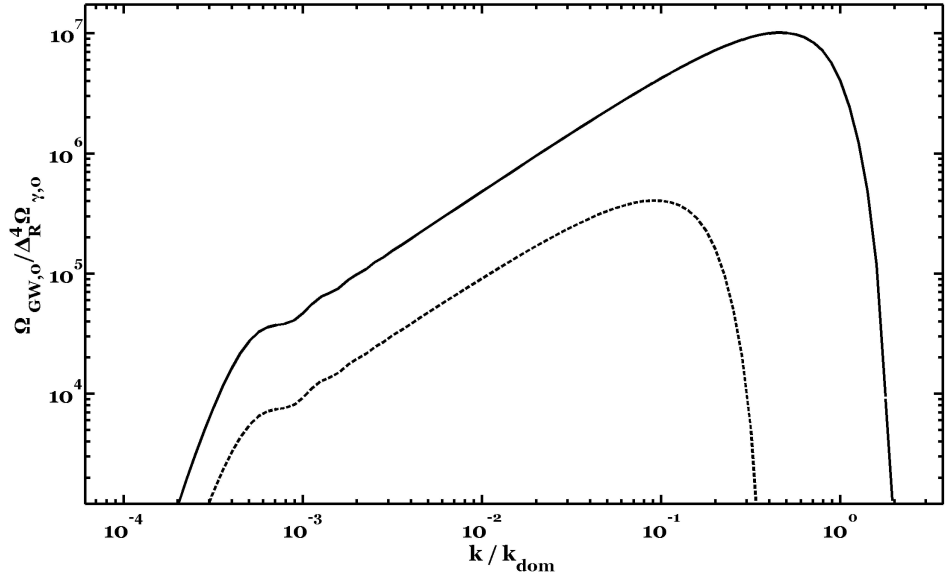} } \caption{The
present energy density of gravitational waves, $\Omega_{GW,0}$,
generated during a matter dominated era shown as a function of
wavenumber $k$. In this example $F=(k_{\rm dom}/k_{\rm
dec})^2=10^6$. The solid line shows the result predicted using the
linear matter power perturbation for $k<k_{\rm dom}$, while the
dotted line shows the result using the matter power spectrum
truncated at $k>k_{\rm cut}=200k_{\rm dec}$.}\label{fig:omega}
\end{figure}

\subsection{Linear scalar perturbations}

If we take Eq.~(\ref{Phdecsub}) for the amplitude of tensor
perturbations for $k<k_{\rm dec}$ at the start of the radiation era,
when $\H=k_{\rm dec}$, we have
\begin{equation}
 \label{GWrad}
 \Omega_{GW}(k,\eta) \simeq \frac{23}{12} \rs \left( \frac{k_{\rm
dom}k}{k_{\rm dec}^2} \right) I_1(k/k_{\rm dom})
  \,,
\end{equation}
and this remains constant (assuming no further production of
gravitational waves on sub-Hubble scales) during the radiation era.
The present day density of second-order gravitational waves produced
due to first-order scalar perturbations is thus given by
\begin{equation}
 \label{OmegaGW0linear}
 \Omega_{GW,0}(k) \simeq \frac{23}{12} \rs
 \Omega_{\gamma,0} \left( \frac{k_{\rm
dom}k}{k_{\rm dec}^2} \right) I_1(k/k_{\rm dom})
  \,,
\end{equation}
The density as a function of wavenumber, $k$, is shown in
Figure~\ref{fig:omega}.

Whereas the power spectrum~(\ref{Phdecsub}) at the end of the early
matter dominated era has a maximum on the Hubble scale at the start
of the radiation era, $k_{\rm dec}$, we find that the present
density of gravitational waves is largest on comoving scales of
order the Hubble size at the start of the early matter era, $k_{\rm
dom}$,
\begin{equation}
 \label{Omegamax}
 \Omega_{GW,0}(k_{\rm dom}) \approx \rs \, \Omega_{\gamma,0} \, \left( \frac{k_{\rm
dom}}{k_{\rm dec}} \right)^2
  \,.
\end{equation}

We find that the maximum density of gravitational waves generated
from linear density perturbations during an early matter dominated
era is enhanced with respect to the naive expectation,
$\Omega_{GW,0}(k_{\rm dom}) \sim \rs \, \Omega_{\gamma,0}\sim
3\times10^{-22}$, by a factor
\begin{equation}
 \label{defF}
F^2 = \left( \frac{k_{\rm dom}}{k_{\rm dec}} \right)^2 \,.
\end{equation}
% $(k_{\rm dom}/k_{\rm dec})^2$ which may be large.
This enhancement factor represents enhanced amplitude with respect
to the gravitational waves that would be produced by scalar metric
perturbations ${\cal P}_h(k_{\rm dom},\eta_{\rm dom}) \sim \rs$ for
modes which re-entered the Hubble scale at the start of the early
matter era but then redshifted once inside the Hubble scale. Instead
we find that the tensor perturbations remain constant even on
sub-Hubble scales while they are supported by constant linear metric
perturbations during the matter era, leading to ${\cal P}_h(k_{\rm
dom},\eta_{\rm dec}) \sim \rs$. The energy density of such modes
during the radiation dominated era is proportional to their
frequency, $k^2$, and we find $\Omega_{GW,0}(k_{\rm dom})\approx F^2
\rs \Omega_{\gamma,0}$.

The enhancement factor calculated by continuing to use the linear
result for scalar metric perturbations, $\dot\Phi_\k=0$, down to
very small scales, $k_{\rm dom}\gg k_{\rm dec}$, could be very large
indeed. It is determined by the duration of the early
matter-dominated era, for which $\H=aH\propto t^{-1/3}\propto
H^{1/3}$, and we have
\begin{equation}
 \label{FH}
F^2 = \left( \frac{H_{\rm dom}}{H_{\rm dec}} \right)^{2/3} \,.
\end{equation}

If the early matter-dominated era arises due to a massive scalar
field displaced from the minimum of its potential in the very early
universe,
%Once the Hubble rate drops below the mass of the field, the field
%will begin to oscillate, and it has the time-averaged
%equation-of-state of non-relativistic, pressureless matter.
then we require $H_{\rm dom}<m$, the mass of the field, and the
matter era will end when the field decays, $H_{\rm dec}\sim\Gamma$,
and thus
\begin{equation}
F^2 < \left( \frac{m}{\Gamma} \right)^{2/3} \,.
\end{equation}

In addition we require that the matter fields decay before
primordial nucleosynthesis and thus $\Gamma>H_{\rm BBN}\sim {\rm
MeV}^2/M_{\rm Pl}$ where $M_{\rm Pl}^2=G^{-1}$. In practice there is
a tighter bound on the Hubble rate at decay for particles with mass
$m>10$~TeV, if we require that their decay rate should not be
suppressed by more that the Planck scale, and thus
$\Gamma>m^3/M_{\rm Pl}^2$. In the optimal case where the field is
precisely $10$~TeV we find the weakest bound
\begin{equation}
 \label{gravFbound}
F^2 < \left( \frac{M_{\rm Pl}}{m} \right)^{4/3} \sim 10^{20} \,.
\end{equation}
The bound is more restrictive for lighter fields, for which $H_{\rm
dom}$ must be smaller, or more massive fields, which must decay
earlier, unless their decay rate is suppressed with respect to
gravitational strength decay. Nonetheless large enhancement factors
would be possible in early matter-dominated eras that are
sufficiently long-lived.

%Note that even if $F^2\sim 10^{20}$ then the maximum power of tensor
%metric perturbations at the start of the radiation era, ${\cal
%P}_h^{\rm max}$ given in Eq.~(\ref{Phmax}), remains of order
%$10^{-7}$. On the other hand the effective density of gravitational
%waves, $\Omega_{GW}$ in Eq.~(\ref{GWrad}), becomes of order unity
%for $k\sim k_{\rm dom}$. Limits on the effective number of light
%particle species at the time of primordial nucleosynthesis require
%$\Omega_{GW,0}^{\rm max}<0.1\Omega_{\gamma,0}$ \cite{Sarkar:1995dd}
%and hence
% \begin{equation}
% \label{Fmax}
% F^2_{\rm max} < 10^{16} \,.
%\end{equation}

\subsection{Non-linear cut-off}

In practice we have seen that the validity of linear results for the
scalar perturbations is restricted to a limited range of scales due
to non-linear effects on scales $k>k_{NL}$ where $k_{NL}$ is given
by Eq.~(\ref{kNL}). Thus the maximum value of $F^2$ for which we can
reliably use the calculation based on linear scalar perturbations is
when $k_{\rm dom}=k_{NL}(\eta_{\rm dec})$ and hence
\begin{equation}
F^2 = \left( \frac{k_{NL}(\eta_{\rm dec})}{k_{\rm dec}} \right)^2
\simeq {\cal P}^{-1/2} \sim 3\times10^4 \,.
\end{equation}

If we assume that non-linear effects give a rapid suppression of the
scalar metric perturbations on scales $k>k_{NL}$ then this
non-linear cut-off suppresses the amplitude of gravitational waves, especially below the cut-off scale.
%
%{}From (\ref{omegap}) we get the fractional energy density of the
%gravitational waves during the radiation era:
%\begin{equation}
% \label{GWrad2}
% \Omega_{GW}(k,\eta) \simeq 4\times\rs \left( \frac{k_{\rm
%cut}^{4}}{k_{\rm dec}^2k^{2}} \right)_{dec}
%\end{equation}
For $k_{NL}<k_{\rm dom}$ we find, from Eqs.~(\ref{Phdecsub3})
and~(\ref{OmegaGW0}), that for $k_{\rm dec}<k<k_{NL}$
\begin{equation}
 \label{NLOmegak}
 \Omega_{GW,0}(k) \simeq 4\rs \, \Omega_{\gamma,0} \, \left( \frac{k_{NL}k}{k_{\rm dec}^2} \right)
 \simeq 5\Delta^{3.5}_{\cal R}\, \Omega_{\gamma,0} \, \left( \frac{k}{k_{\rm dec}} \right)  \,.
\end{equation}
The maximum value of the energy density of gravitational waves at
the present time is reached for $k\simeq k_{NL}$:
\begin{equation}
\label{OmegamaxNL}
 \Omega_{GW,0}(k_{NL}) \simeq 4\rs \left( \frac{k_{NL}}{k_{\rm dec}}
\right)^2\sim 6\triangle_{\cal{R}}^{3} \Omega_{\gamma,0}
% \sim 10^{-17}
  \,.
\end{equation}
For primordial density perturbations with the same amplitude on
small scales as seen on CMB scales today,
$\triangle_{\cal{R}}^{2}\simeq 2\times 10^{-9}$, this gives
$\Omega_{GW,0}(k_{NL})\sim 10^{-17}$.
This result neglects any source for the gravitational waves coming
from non-linear scales, $k>k_{NL}$, and thus gives a very
conservative lower bound on the density of gravitational waves
produced during an early matter era.

\section{Detectability}

\subsection{Frequency ranges}

The comoving Hubble scale during a radiation-dominated era is given
by $k=aH$ where
\begin{equation}
H^2 = \frac{8\pi G}{3}\, \left( g_* \, \frac{\pi^2}{30} \, T^4
\right) \,.
\end{equation}
where $g_*$ is the effective number of degrees of freedom at
temperature $T$.

At the present time this corresponds to a physical scale
\begin{equation}
\lambda_0 = \frac{2\pi a_0}{k} \approx 2\times 10^{16}\, g_*^{-1/6}
\left( \frac{{\rm GeV}}{T} \right)\, {\rm m} \,,
\end{equation}
and a frequency
\begin{equation}
 \label{fT}
\f = \frac{c}{\lambda_0} \approx 1.2 \times 10^{-8}\, g_*^{1/6}
\left( \frac{T}{{\rm GeV}} \right)\, {\rm Hz} \,.
\end{equation}

The standard radiation dominated hot big bang must be restored
before the epoch of primordial nucleosynthesis, and thus the
temperature after the decay of an early matter-dominated era,
$T_{\rm dec}$, must be above $1$~MeV and thus
\begin{equation}
\f_{\rm dec} > 10^{-11}\ {\rm Hz} \,.
\end{equation}
Note that ratio between the present frequency of Hubble-scale modes
at the start of the matter era and Hubble-scale modes at the end of
the matter era corresponds to the enhancement factor in
Eq.~(\ref{defF})
\begin{equation}
F = \frac{\f_{\rm dom}}{\f_{\rm dec}} \,.
\end{equation}
Thus a large enhancement factor, $F$, also implies a wide range of
frequencies over which gravitational waves are generated during an
early matter era.

Successful, standard, big-bang nucleosynthesis (BBN)
\cite{Sarkar:1995dd} places an upper bound on the density of
primordial gravitational waves,
 \begin{equation}
 \label{BBN}
\Omega_{GW,0}<0.1\Omega_{\gamma,0}\simeq 10^{-6}
 \,,
\end{equation}
on any scales smaller than the Hubble scale at that time,
corresponding to $\f>\f_{BBN}\simeq 10^{-11}$~Hz.
Gravitational waves generated during a early matter era are only
present on LIGO scales, $f_{\rm LIGO}\sim 100$~Hz, if $T_{\rm
dec}<10^{10}$~GeV, and on LISA scales, $\f_{\rm LISA}\sim
10^{-3}$~Hz, if $T_{\rm dec}<10^5$~GeV.
%, where we identify $T_{\rm dom}$ with the equivalent temperature at
% the start of the early matter era
%  \begin{equation}
%  T_{\rm dom} = \left( \frac{H_{\rm dom}}{H_{\rm dec}} \right)^{1/2} T_{\rm dec}
%  = F^{3/2}  T_{\rm dec} \,.
%  \end{equation}
%
Current limits from the Laser Interferometer Gravitational-wave
Observatory (LIGO) bound $\Omega_{GW,0}<6\times10^{-5}$ in the
frequency range $\f_{\rm LIGO}\sim100$~Hz with Advanced LIGO
sensitive to $\Omega_{\rm GW,0}\sim 10^{-9}$ in the future
\cite{Abbott:2006zx}.
LISA should be able to detect a primordial GW background
$\Omega_{GW,0}\sim10^{-11}$ at frequencies $\f_{\rm
LISA}\sim10^{-3}$~Hz \cite{Hogan:2006va}, and future experiments
such as Big Bang Observer (BBO) may be able to detect a primordial
background $\Omega_{GW,0}\sim10^{-17}$ at frequencies $\f_{\rm
BBO}\sim1$~Hz \cite{Corbin:2005ny}.

\subsection{Reheating after GUT-scale inflation}
\label{GUTinf}

Most inflationary models of the early universe incorporate an early
matter dominated period immediately after inflation when the energy
density of the universe is dominated by oscillating, massive scalar
fields. If the fields decay perturbatively to radiation this
corresponds to a period of reheating.
Current upper bounds from the CMB on first-order gravitational waves
produced from vacuum fluctuations during inflation
\cite{Komatsu:2008hk} place a bound on the maximum value of the
energy density during inflation, $\rho_{\rm inf}^{1/4}=M_{\rm
inf}<10^{16}$~GeV, and thus the Hubble scale at the start of the
early matter era.
% and hence $T_{\rm dom}<10^{16}$~GeV, corresponding to $f_{\rm
% dom}<10^{11}$~Hz, well above LISA or LIGO frequencies.

Models such as chaotic inflation driven by a massive scalar field
come close to saturating this bound and thus correspond to GUT-scale
inflation.
On the other hand constraints from the thermal production of
gravitinos in supersymmetric models suggest that the maximum reheat
temperature, $T_{\rm dec}$, should be less than about $10^9$~GeV
\cite{Sarkar:1995dd}\footnote{Note that in the case of reheating at
the end of chaotic inflation, the mass of the inflaton is required
to be $m\sim 10^{13}$~GeV, and delaying reheating so that $T_{\rm
dec}<10^9$~GeV then actually violates the bound given in
Eq.~(\ref{gravFbound}) for this mass.}. This implies, from
Eq.~(\ref{FH}), a lower bound on the enhancement factor, $F^2>10^8$,
suggesting that the power spectrum of second-order gravitational
waves generated during reheating after GUT-scale inflation could be
larger than that from first-order gravitational waves at frequencies
$\f \sim \f_{\rm dom}$, and possibly much larger.

Taking Eq.~(\ref{ns}) for linear density perturbations on all scales
$k<k_{\rm dom}$, we have, from Eqs.~(\ref{Omegamax}) and~(\ref{FH}),
\begin{eqnarray}
 \label{chaoticOmegadom}
 \Omega_{GW,0} (\f_{\rm dom}) &\approx& 3\times10^9 g_*^{-1/3} \rs
 \Omega_{\gamma,0} \left( \frac{M_{\rm inf}}{10^{16}~{\rm GeV}}
 \right)^{4/3} \left( \frac{T_{\rm dec}}{10^9~{\rm GeV}}
 \right)^{-4/3} \,,\\
 &\approx& 10^{-12} g_*^{-1/3} \left( \frac{M_{\rm inf}}{10^{16}~{\rm GeV}}
 \right)^{4/3} \left( \frac{T_{\rm dec}}{10^9~{\rm GeV}}
 \right)^{-4/3} \,,
\end{eqnarray}
where in the second line we set $\Delta_{\cal R}^2\simeq
2.4\times10^{-9}$.
In this case the nucleosynthesis limit on the density of primordial
gravitational waves (\ref{BBN}) places a lower limit on the minimum
allowed reheating temperature, $T_{\rm dec}>10^3$~GeV, assuming
$M_{\rm inf}\sim 10^{16}$~GeV.
This value of $\Omega_{GW,0}$ is obtained at high frequencies
\begin{equation}
 \label{fdom}
\f_{\rm dom}
% = F f_{\rm dec}
 \approx 7\times 10^5~{\rm Hz}~\left(
\frac{T_{\rm dec}}{10^9~{\rm GeV}} \right)^{1/3} \left( \frac{M_{\rm
inf}}{10^{16}~{\rm GeV}} \right)^{2/3} \,,
\end{equation}
which is always above the range of detectors such as LIGO.
%
%, and such large enhancement factors can only be achieved by
%extrapolating the linear matter power spectrum far beyond the
%non-linear scale, $k_{NL}$ defined in Eq.(\ref{kNL}).

At the lower end of the range of generated frequencies, we have from
Eq.~(\ref{fT}) that $\f_{\rm dec}\sim 10(T_{\rm dec}/10^9~{\rm
GeV})$~Hz, which is always below the LIGO range if $T_{\rm
dec}<10^{10}$~GeV. Using the linear perturbations~(\ref{ns}) for all
$k<k_{\rm dom}$, the density of gravitational waves at LIGO
frequencies is then, from Eq.~(\ref{OmegaGW0}),
\begin{eqnarray}
 \Omega_{GW,0}(\f_{\rm LIGO}) &\approx& 5\times10^5 g_*^{-1/3} \rs \Omega_{\gamma,0} \left( \frac{T_{\rm
dec}}{10^9~{\rm GeV}} \right)^{-5/3} \left( \frac{M_{\rm
inf}}{10^{16}~{\rm GeV}} \right)^{2/3} \left( \frac{\f_{\rm
LIGO}}{100 {\rm Hz}} \right)
% \,,
% \\
% &\approx& 10^{-17} \left( \frac{T_{\rm dec}}{10^9~{\rm GeV}}
% \right)^{-5/3} \left( \frac{M_{\rm inf}}{10^{16}~{\rm GeV}}
% \right)^{2/3} \left( \frac{f_{\rm LIGO}}{100 {\rm Hz}} \right)
 \,.
\end{eqnarray}
Taking $\Delta_{\cal R}^2\simeq 2\times10^{-9}$, this would be
within the sensitivity of Advanced LIGO if the reheat temperature
takes a low value, $T_{\rm dec}< 10^5$~GeV:
\begin{eqnarray}
 \label{chaoticOmegaLIGO}
 \Omega_{GW,0}(\f_{\rm LIGO}) &\sim& 10^{-9} \left( \frac{T_{\rm dec}}{10^5~{\rm GeV}}
\right)^{-5/3} \left( \frac{M_{\rm inf}}{10^{16}~{\rm GeV}}
\right)^{2/3} \left( \frac{\f_{\rm LIGO}}{100 {\rm Hz}} \right) \,,
\end{eqnarray}

However Eqs.~(\ref{chaoticOmegadom}--\ref{chaoticOmegaLIGO}) assume
large enhancement factors, extrapolating the linear power spectrum
(\ref{ns}) well into the non-linear regime on small scales, $k_{\rm
dom}\ll k_{NL}$ defined in Eq.(\ref{kNL}). If we assume non-linear
effects cut-off the matter power spectrum below $k_{NL}\simeq200
k_{\rm dec}$ for $\Delta_{\cal R}^2\simeq10^{-9}$, then from
Eq.~(\ref{NLOmegak}) we have for $\f_{\rm dec}<\f_{\rm LIGO}<\f_{NL}$
 \begin{eqnarray}
 \Omega_{GW,0}(\f_{\rm LIGO}) &\approx& 50 g_*^{-1/6} \Delta_{\cal R}^{3.5} \Omega_{\gamma,0} \left( \frac{T_{\rm
dec}}{10^9~{\rm GeV}} \right)^{-1} \left( \frac{\f_{\rm LIGO}}{100
{\rm Hz}} \right) \,\\
 &\approx& 5\times10^{-19}
 \left( \frac{T_{\rm dec}}{10^9~{\rm GeV}} \right)^{-1}
 \left( \frac{\f_{\rm LIGO}}{100 {\rm Hz}} \right) \, .
 \end{eqnarray}
The amplitude increases for lower $T_{\rm dec}<10^9$~GeV but it
reaches a maximum value, $\Omega_{GW,0} \sim 10^{-17}$, at
\begin{equation}
 \label{fNL}
\f_{NL} \approx 2\times10^3~{\rm Hz}~\left( \frac{T_{\rm
dec}}{10^9~{\rm GeV}} \right) \,,
\end{equation}
and this drops below the LIGO range of frequencies for $T_{\rm
dec}<5\times10^7$~GeV.
Advanced LIGO would thus only be able to detect a primordial
gravitational wave background generated by density perturbations in
the linear regime during reheating if the density perturbation
$\Delta_{\cal R}^2 \sim 10^{-3}$ and $T_{\rm dec}\sim10^{10}$~GeV.
However future experiments such as BBO could detect primordial
gravitational waves generated during reheating even for
$\Delta_{\cal R}^2\sim10^{-9}$ if $T_{\rm dec}\sim10^6$~GeV.

\subsection{Reheating after intermediate-scale inflation}
\label{intinf}

In many models of inflation the energy scale of inflation is set by
the intermediate scale $\rho_{\rm inf}^{1/4}=M_{\rm inf} \simeq
10^{10}$~GeV \cite{Lyth:1998xn}. Such models include hybrid
inflation \cite{Linde:1991km,Linde:1993cn,Copeland:1994vg} in which
case the inflaton mass and Hubble-scale at the end of inflation is
of order TeV. Tachyonic instability leads to the end of inflation
when the inflaton reaches a critical value and the vacuum energy
density driving inflation is rapidly converted to oscillating scalar
fields. The spatially coherent oscillations of the inflaton fields
rapidly fragment \cite{Felder:2000hj}, but if a significant fraction
of the energy density goes into weakly coupled massive fields which
decay perturbatively then we may have an extended period of
reheating after inflation.

The lower energy scale during inflation, compared with GUT-scale
inflation, leads to a lower frequency of gravitational waves
produced at the end of inflation. {}From Eq.~(\ref{fdom}) we have
\begin{equation}
\f_{\rm dom} \approx 0.7~{\rm Hz}~\left( \frac{T_{\rm dec}}{10^3~{\rm
GeV}} \right)^{1/3} \left( \frac{M_{\rm inf}}{10^{10}{\rm GeV}}
\right)^{2/3} \,,
\end{equation}
If $T_{\rm dec}<10^9$~GeV then this is below the frequency range for
LIGO but could be detectable by BBO if $T_{\rm dec}\sim10^3$~GeV.
% Lower frequency GWs generated during reheating would be in the LISA
% frequency range if $T_{\rm dec}<1$~keV.
Indeed if the inflaton or other moduli field excited at the end of
inflation are weakly coupled and decay only just before big bang
nucleosynthesis, $T_{\rm dec}\sim 1$~MeV, we have $\f_{\rm dom}\sim
\f_{\rm LISA}$ and a maximal enhancement factor $F^2\sim10^{17}$ for
$M_{\rm inf}\sim 10^{10}$~GeV, which would be large enough to
violate the BBN bound in Eq.~(\ref{BBN}).

Such large enhancement factors imply that the small scale density
perturbations are far into the non-linear regime. If non-linear
evolution suppresses the scalar perturbations on scales $k>k_{NL}$
defined in Eq.~(\ref{kNL}), then the gravitational waves generated
are suppressed for $\f>\f_{NL}$ where
\begin{equation}
\f_{NL} \approx 2\times10^{-3}~{\rm Hz}~\left( \frac{T_{\rm
dec}}{10^3~{\rm GeV}} \right) \,,
\end{equation}
and we have set $\Delta_{\cal R}^2\approx 2.4\times10^{-9}$. Thus we
find $\nu_{NL}<\nu_{\rm LISA}$ for $T_{\rm dec}<10^3$~GeV.  For
$T_{\rm dec}\sim 10^4$~GeV, LISA could detect this background if
$\Delta_{\cal R}^2\approx 10^{-5}$ on these scales.

\subsection{Moduli or curvaton domination}

Any scalar field, with mass $m$, displaced from the minimum of its effective potential after inflation begins to oscillate with an initial amplitude $\chi_{\rm osc}$ either at the end of inflation, when the Hubble rate is $H_{\rm osc}\simeq H_{\rm inf}$ if $m>H_{\rm inf}$, or once the Hubble rate drops below the mass, $H_{\rm osc}\simeq m$, if $m<H_{\rm inf}$. The fractional energy density of the field at this time is
\begin{equation}
\Omega_{\chi,{\rm osc}}
 = \frac{4\pi}{3} \left( \frac{m}{H_{\rm osc}} \right)^2
 \left( \frac{\chi_{\rm osc}}{M_{\rm Pl}} \right)^2 \,.
\end{equation}
Note that $\Omega_{\chi,{\rm osc}}<1$ since $\chi_{\rm osc}<M_{\rm
Pl}$ , otherwise the field would slow-roll and drive a period of
inflation. During a radiation dominated era following inflation, the
density of the oscillating field grows relative to the radiation,
eventually coming to dominate the energy density and drive an early
matter era when the Hubble rate drops below \cite{Dimopoulos:2004yb}
\begin{equation}
 H_{\rm dom} \simeq \Omega_{\chi,{\rm osc}}^2 H_{\rm osc}
  \simeq
 \left( \frac{4\pi}{3} \right)^2 \left( \frac{m}{H_{\rm osc}} \right)^3
 \left( \frac{\chi_{\rm osc}}{M_{\rm Pl}} \right)^4 m \,.
\end{equation}
The duration of the early matter era depends on the decay time of the field, $H_{\rm dec}=\Gamma$, which must be before primordial nucleosynthesis. The duration is described by the factor $F^2$ defined in Eq.~(\ref{FH})
\begin{equation}
 \label{moduliF}
F^2 \simeq 10^{20} g_*^{-1/3}
 \left( \frac{\chi_{\rm osc}}{M_{\rm Pl}} \right)^{8/3}
 \left( \frac{m}{H_{\rm osc}} \right)^2
 \left( \frac{m}{10~{\rm TeV}} \right)^{2/3} \left( \frac{T_{\rm dec}}{{\rm MeV}} \right)^{-4/3} \,,
\end{equation}
which can be very large if $\chi_{\rm osc}\sim M_{\rm Pl}$, becoming
largest for fields which decay just before nucleosynthesis, $T_{\rm
dec}\sim 1$~MeV. Equation~(\ref{moduliF}) is consistent with
Eq.~(\ref{chaoticOmegadom}) for the amplitude of gravitational waves
produced from reheating at the end of inflation if we set $\chi_{\rm
osc}\sim M_{\rm Pl}$ and $H_{\rm osc}\sim m\sim H_{\rm inf}$.

If a moduli field is light during inflation, $m<H_{\rm inf}$, then
it acquires a spectrum of perturbations on large, super-Hubble
scales, from initial vacuum fluctuations on small, sub-Hubble
scales. In the curvaton scenario
\cite{Linde:1996gt,Enqvist:2001zp,Lyth:2001nq,Moroi:2001ct} these
inhomogeneities about the background expectation value of the field
is supposed to give rise to the primordial density perturbation, and
hence in the simplest curvaton model in which the curvaton comes to
dominate the total energy density of the universe before decaying,
observations require \cite{Lyth:2001nq}
\begin{equation}
\Delta_{\cal R}^2 \simeq \left( \frac{H_{\rm inf}}{3\pi\chi_{\rm osc}} \right)^2 \,.
\end{equation}
This determines the initial amplitude of the curvaton oscillations
in terms of the inflationary energy scale, $\chi_{\rm osc} \sim
\Delta_{\cal R}^{-1} H_{\rm inf}$, and hence from
Eqs.~(\ref{Omegamax}) and~(\ref{moduliF}) we can calculate the peak
amplitude of gravitational waves generated from linear density
perturbations
\begin{equation}
 \label{curvatonF}
%F^2 \simeq 2\times10^8 g_*^{-1/3}
% \left( \frac{H_{\rm inf}}{10^7~{\rm GeV}} \right)^{8/3}
% \left( \frac{m}{10~{\rm TeV}} \right)^{2/3} \left( \frac{T_{\rm dec}}{{\rm MeV}} \right)^{-4/3} \,,
\Omega_{GW,0}(\f_{\rm dom}) \simeq 6\times10^{-14} g_*^{-1/3}
 \left( \frac{H_{\rm inf}}{10^7~{\rm GeV}} \right)^{8/3}
 \left( \frac{m}{10~{\rm TeV}} \right)^{2/3} \left( \frac{T_{\rm dec}}{{\rm MeV}} \right)^{-4/3} \,,
\end{equation}
where we have taken $\Delta_{\cal R}^2\simeq 2.4\times10^{-9}$ and
$m=H_{\rm osc}<H_{\rm inf}$, and we note that there is a lower bound
on the Hubble rate during inflation, $H_{\rm inf}>10^7$~GeV
\cite{Lyth:2003dt} in the simplest curvaton models. Note also that
if the curvaton comes to dominate the energy density of the universe
before it decays, the primordial density perturbation is close to
Gaussian (the non-linearity parameter $f_{\rm NL}\simeq -5/4$
\cite{Bartolo:2003jx,Lyth:2005du,Malik:2006pm,Sasaki:2006kq,Assadullahi:2007uw}).

Thus we see that an early matter dominated era may have a long
duration in the curvaton scenario leading to a large enhancement
factor in our calculation based on the linear density perturbation.
However, as in the case of reheating after intermediate-scale
inflation, the frequency is below LIGO scales if the decay
temperature is low. The highest frequency GW generated in the
curvaton-dominated era would be at frequency
\begin{equation}
\f_{\rm dom} = F \f_{\rm dec} \simeq 10^{-7} \left( \frac{H_{\rm
inf}}{10^7~{\rm GeV}} \right)^{4/3} \left( \frac{m}{10\ {\rm TeV}}
\right)^{1/3}\left( \frac{T_{\rm dec}}{{\rm MeV}} \right)^{1/3} \
{\rm Hz}
 \,,
 \end{equation}
where the lowest frequencies would be, from Eq.~(\ref{fT}),
\begin{equation}
\f_{\rm dec} \simeq 10^{-11} \left( \frac{T_{\rm dec}}{{\rm MeV}}
\right) \ {\rm Hz}
 \,.
 \end{equation}

If non-linearities suppress the density perturbation on sub-Hubble
scales in an early matter era then the only gravitational waves that
are not reshifted away will be on the lowest frequencies,
$\f<\f_{\rm NL}$ and we recover the same results as in the case of
reheating after inflation, given in sub-sections~(\ref{GUTinf})
and~(\ref{intinf}) determined by the final decay temperature,
$T_{\rm dec}$.

\section{Conclusions}

In a matter-dominated era the behaviour of the linearised density
perturbations is particularly simple. The gravitational instability
of pressureless matter leads to the growth of the comoving density 
perturbation, while the Newtonian metric potential, $\Phi$, remains
constant on all scales. This provides a constant source term,
$S_{\bf k}$, at second-order in the wave equation for tensor metric
perturbations, leading to the general solution given in Eq.~(\ref{h1}).
At late times this approaches the constant solution, given in
Eq.~(\ref{h3 small scale}), on sub-Hubble scales.

This constant tensor mode during the matter era is quite different
from the behaviour we would expect for a free gravitational wave on
sub-Hubble scales.
% In addition we must be aware the tensor metric perturbation at
% second order is gauge dependent \cite{Bruni,Malik}.
Indeed the free part of the tensor perturbation, given by the
oscillating terms in Eq.~(\ref{h1}), are redshifted away
$|h_\k|\propto a^{-1}$, diluting any pre-existing gravitational
waves on sub-Hubble scales at the start of a matter-era. But the
constant amplitude of tensor metric perturbations in the Poisson
gauge, supported by first-order metric perturbations in an early
matter era, $h_\k\propto S_\k/k^2$, becomes the initial condition
for the amplitude of freely oscillating gravitational waves in the
subsequent radiation era when $\Phi$ rapidly decays on sub-Hubble
scales.

We have seen that a constant amplitude of scalar metric
perturbations, ${\cal P}_\Phi\sim \Delta_{\cal R}^2$, extending down
to sub-Hubble scales can lead to a large enhancement in the
amplitude of gravitational waves. On the smallest scales which we
considered, $k_{\rm dom}$ corresponding to the comoving Hubble scale
at the start of matter domination, the enhancement factor is easily
understood due to the amplitude of tensor perturbations on this
scale remaining constant during the matter dominated era instead of
being redshifted on sub-Hubble scales. As a result the energy
density of gravitational waves in the late universe is enhanced with
respect to the naive estimate $\Omega_{GW,0}\sim
\rs\Omega_{\gamma,0}$, by a factor $F^2=(k_{\rm dom}/k_{\rm
dec})^2$, where $k_{\rm dec}$ is the comoving Hubble scale at the
end of the matter era.
For $F^2>r \Delta_{\cal R}^{-2}\sim 4\times 10^8 r$, where $r\ll 1$
is the usual tensor-scalar ratio at first-order
\cite{Komatsu:2008hk}, the second-order gravitational waves
generated from linear perturbations during an early matter dominated
era would dominate over those produced directly from vacuum
fluctuations in the gravitational field during inflation.
Large enhancement factors occur when there is a sufficiently long
matter dominated era in the early universe, preceding the standard
radiation dominated era required for successful primordial
nucleosynthesis. This is certainly possible given our state of
ignorance of the thermal history of the universe before
nucleosynthesis. The spectrum of gravitational waves could provide
one of the few constraints on the universe before nucleosynthesis.

However, a sufficiently long-lived matter era also predicts a
breakdown of linear perturbation theory for density perturbations on
sub-Hubble scales. Predictions based on linear theory can only be
trusted for $F^2<(k_{NL}/k_{\rm dec})^2\sim \Delta_{\cal R}^{-1}$. A
constant scalar metric perturbation in linear theory in the
longitudinal gauge, $\Phi$, implies a growing comoving density
perturbation, $\delta\rho/\rho$. If non-linear evolution of the
matter perturbations suppresses the scalar source term on scales
$k>k_{NL}$ then the peak in the gravitational wave density is
shifted from $k_{\rm dom}$ to $k_{NL}$ and the maximum amplitude at
this peak is reduced to $\Omega_{GW,0}\sim 6\Delta_{\cal
R}^3\Omega_{\gamma,0}\sim 10^{-17}$ for $\Delta_{\cal R}^2 \simeq 2\times10^{-9}$ 
which would only be detectable
by experiments in the far future such as the Big Bang Observer.

In summary we have seen that growing density perturbations during an
early matter era could lead to a detectable background of primordial
gravitational waves. But this is only detectable with planned
experiments such as LIGO and LISA if the primordial metric
perturbations are significantly larger on small scales 
($\Delta_{\cal R}^2 \sim 6\times10^{-4}$ for LIGO, 
or $\Delta_{\cal R}^2 \sim 3\times10^{-5}$ for LISA) 
than currently observed on very large scales, where $\Delta_{\cal R}^2\simeq
2\times10^{-9}$, or if significant non-linear density perturbations survive
on sub-Hubble scales during the early matter era. This latter case
requires non-linear (and hence probably numerical) solutions for the
density perturbations. Because these perturbations are on sub-Hubble
scales, Newtonian theory may be sufficient on these scales. There
has already been investigation of gravitational waves generated by
nonlinear density perturbations on sub-Hubble scales during
preheating after inflation
\cite{Khlebnikov:1997di,Easther:2006gt,Easther:2006vd,Dufaux:2007pt,GarciaBellido:2007af,GarciaBellido:2007dg,Dufaux:2008dn},
and our work suggests that similar numerical calculations will also
be required to investigate whether a detectable level of
gravitational waves could be produced from gravitational instability
in an early matter era.

\section*{Acknowledgements}

HA is supported by the British Council. DW is supported by the STFC.
The authors are grateful to Rob Crittenden and Kazuya Koyama for
useful discussions.

%\bibliographystyle{utphys}
%\bibliography{bibtex2}

\providecommand{\href}[2]{#2}\begingroup\raggedright\endgroup

\end{document}